\documentclass[reprint,eqsecnum,floats,aps,amsmath,amssymb,nofootinbib,prd,onecolumn, showpacs]{revtex4-2}

\usepackage{bm}
\usepackage{graphicx,physics}
\usepackage{graphicx}
\usepackage{amsmath,amssymb,mathtools,mathrsfs}
\usepackage{amsmath,amssymb,mathrsfs}
\usepackage{hyperref}
\usepackage{graphicx}
\usepackage{arydshln}
\usepackage{xcolor}
\usepackage{braket}
\usepackage{tensor}
\usepackage{enumitem,array,textcomp}
\usepackage[scr=boondoxo]{mathalpha}
\usepackage{relsize}

\usepackage{amsmath}
\usepackage{accents} 
\usepackage{orcidlink}

\usepackage{color}

\begin{document}

\title{Unitary implementability of Darboux transformations}

\author{Guillermo  A. Mena Marugán\,\orcidlink{0000-0003-3378-9610}}
\email{Electronic address: mena@iem.cfmac.csic.es}
\affiliation{Instituto de Estructura de la Materia, IEM-CSIC, C/ Serrano 121, 28006 Madrid, Spain}

\author{Alvaro Torres-Caballeros\,\orcidlink{0000-0003-0108-1757}}
\email{Electronic address: alvaro.torres@iem.cfmac.csic.es}
\affiliation{Instituto de Estructura de la Materia, IEM-CSIC, C/ Serrano 121, 28006 Madrid, Spain}

\author{José Velhinho\,\orcidlink{0000-0002-3219-903X}}
\email{Electronic address: jvelhi@ubi.pt}
\affiliation{Faculdade de Ci\^encias and FibEnTech-UBI, Universidade da Beira Interior, R. Marqu\^es D'Ávila e Bolama, 6201-001 Covilh\~a, Portugal}

\def\reps{representations\ }
\def\l{\hat{l}}
\def\bl{\Tilde{l}}
\def\bll{\hat{\Tilde{l}}}
\def\bk{\Tilde{k}}
\def\z{z_{\bk\bll}}
\def\fhi{\tilde{\phi}_{n l m}}
\def\p{\mathrm{ax}}
\def\a{\mathring{a}_{n l m}}
\def\aa{\mathring{a}_{n l m}^{*}}
\def\S{\mathscr{S}_{\l}}
\def\SS{\tilde{\mathscr{S}}_{\l}}
\def\SSS{\tilde{\mathfrak{s}}_{\bll}}
\def\n{\mathfrak{n}}
\newcommand{\dtilde}[1]{\tilde{\raisebox{0pt}[0.85\height]{$\tilde{#1}$}}}
\def\DS{\mathscr{\Tilde{S}}_{\frac{n}{l}}}
\def\b{b_{\l}}

\def\x{\xi_{nl}}
\def\f{f_{nl}}
\def\g{g_{nl}}
\def\vaf{\vartheta_{nl}^{(f)}}
\def\vag{\vartheta_{nl}^{(g)}}
\def\s{s_{\l}}
\def\h{h_{nl}}
\def\bt{\Tilde{b}_{\l}}
\def\Hb{\mathscr{H}}
\def\bN{\mathrm{N}}
\def\GS{\mathrm{GS}}
\def\M{\mathrm{CPM}}
\def\MM{\textstyle\widetilde{\scriptstyle \mathrm{CPM}}}
\def\GGSS{\textstyle\widetilde{\scriptstyle \mathrm{GS}}}

\def\dt{Darboux transformation}
\def\dts{Darboux transformations}

\def\bs{\accentset{\mathlarger{\star}}{b}_{\l}}

\begin{abstract}
We investigate the unitary implementability of the canonical transformations corresponding to Darboux transformations in the Hamiltonian formalism recently developed for gauge invariant perturbations of nonrotating black holes. Starting from the formalism in the black hole interior, we demonstrate that these canonical transformations are unitarily implementable under general conditions on the effective potentials appearing in the master equations of the perturbations. We also show that the privileged massless Fock representation of the perturbative variables selected by admitting (Heisenberg) unitary dynamics, employed in our quantization, is not unitarily equivalent to the simple massless representation of the master variables satisfying the effective Regge-Wheeler equation. This highlights the importance of the choice of master variables in which the quantum representation is initially defined. 
Furthermore, we show that not all such choices of variables possessing unitarily implementable dynamics can be reached through Darboux transformations, but rather through more general canonical transformations. These results advance our understanding of the relationship between well-established classical relations in black hole perturbation theory and their quantum representations, explaining how isospectral Darboux transformations can be extended to include all the infinitely many possible values of the angular momentum such that they are unitarily implementable in a quantum field theory context.
\end{abstract}

\maketitle

\section{Introduction}

Black hole (BH) perturbation theory offers a privileged and natural arena for exploring a vast range of phenomena, encompassing stability analyses to understand the fate of classical singularities (see e.g. Ref. \cite{Singularity}), response dynamics to small disturbances like infalling matter, nearby objects, or binary mergers, and gravitational wave observations, which bring BH dynamics into direct contact with experiment \cite{Nollert,Berti,LISA,ET,Tests,Barack}.  Additionally, within this set of fertile fields to explore, one finds the compelling challenge of passing from classical to quantum descriptions of gravity \cite{Frolov,TThiemann}. BHs are by themselves a canonical scenario to probe strong gravity regimes, where quantum geometry phenomena are expected to be relevant. Besides, given the nonlinear character of General Relativity, the separation of the background geometry and small fluctuations around it provides a more tractable setting than the full theory for investigating how gravitational degrees of freedom can be represented and quantized. Although this obviously does not solve the problem of quantum gravity, it isolates fundamental conceptual questions that can be addressed within a well-defined framework. 

On the other hand, it is well-known that if we describe perturbations of the geometry by metric perturbations, some of those perturbations correspond in fact to perturbative gauge degrees of freedom. Since the background spacetime and the perturbed one are distinct manifolds, one introduces a map to identify points on both manifolds and uses its pullback to compare the physical, perturbed metric with the background one. However, there does not exist a unique or privileged identification, which yields different expressions for the metric perturbations. Hence, an infinitesimal change of the identification map characterized by a vector field will result in a Lie derivative correction term of the first-order metric perturbation, that is essentially as \textit{small} as the former, i.e. of the same perturbative order. Indeed, we can circumvent this issue by choosing a particular gauge and fixing this freedom \cite{ReggeWheeler}. But to prevent us from losing generality and to ensure that both our observables (once we move to the quantum realm) and our results do not depend on our choice, the most satisfactory approach is to work with perturbative quantities that are gauge invariants \cite{ReggeWheeler,Bardeen,Mukhanov,Sasaki}. They can be constructed by summing background-dependent combinations of the linear perturbative terms, and imposing that they have vanishing Poisson brackets with the relevant perturbative gauge constraints (at least weakly), obtained in turn by linearization of the perturbed diffeomorphism constraints of General Relativity\footnote{In fact, these perturbative gauge constraints can be Abelianized (at the considered dominant perturbative order), simplifying all considerations \cite{Master,GRG}.} \cite{Gpap,Master}.

Even though gauge freedom can be controlled in this way, the true physical degrees of freedom of the perturbations can still admit different descriptions in terms of different variables. Employing expansions in spherical harmonics, these gauge invariant variables can be identified with the coefficients of the expansions, and therefore be treated as fields in $1+1$ dimensions \cite{GRG} (with the dimensions of the spherical orbits of symmetry removed). These fields are usually called master functions, or master variables, and their dynamics are governed by evolution equations, called master equations, that can be written as (bidimensional) wave equations subject to a potential. Different choices of master functions imply distinct potentials. The Regge-Wheeler and Zerilli equations, respectively governing the Cunningham-Price-Moncrief (CPM) and Zerilli-Moncrief master functions, usually adopted to describe the axial and polar gravitational perturbations of the Schwarzschild spacetime respectively, serve as an illustrative example \cite{ReggeWheeler,CPM,Ze,Mo}. Remarkably, although their potentials differ, the two equations (and the respective master functions satisfying them) are related by a Darboux transformation (DT) \cite{Darboux}. Moreover, the existence of this transformation ultimately impacts the associated spectrum of quasinormal modes \cite{Chandra}, namely the complex-frequency solutions to the master equations such that they satisfy exterior boundary conditions of ingoing nature at the BH horizon and outgoing nature at spatial infinity \cite{Nollert}. Indeed, the DT ensures that the two related master equations present the same quasinormal frequencies when one considers any fixed value of the angular momentum in the spherical harmonics decomposition \cite{Chandra1,Chandra2,Chandra3,Yurov,Glampe}. This property is referred to as isospectrality. An interesting avenue of research to investigate is whether this isospectrality persists when the classical background is replaced with an effective geometry that incorporates quantum corrections. Recent studies indicate that isospectrality may be broken within certain effective loop quantum BH models \cite{Leno,CorralOlmedo}, such as the widely studied Ashtekar-Olmedo-Singh model \cite{AOS,AOS2,AO}, if the potentials of the master equations are evaluated directly on the effective background without any further effective modifications. This calls for a deeper analysis, to understand whether the results are a mere artifact of the simple prescription used to implement quantum corrections effectively, or whether the reason for isospectrality breaking can be traced to a more fundamental, intrinsic feature of the quantization of the perturbations in the modified effective background geometry.

Traditionally formulated at the level of master functions in a second-order formalism (like in Lagrangian formulations), DTs for BHs constitute maps from one differential equation governing the behavior of the perturbations, in principle in the BH exterior, to another equation of the same general type. This is achieved by transforming the perturbative variables and, consequently, the potential of its associated equation of motion, while establishing a correspondence between the solutions of the original and transformed equations, which reflects a subjacent mathematical symmetry between the two apparently different descriptions \cite{Chandra,CM1,CM2}. Hence, while two differential (master) equations seem different, their solutions can encode equivalent physical dynamics. However, it is important to mention that not every pair of different potentials is related by a DT; the potentials and transformation must satisfy specific consistency conditions, including a particular Riccati equation for the function defining the transformation, as we will later specify. Conversely, DTs can also be understood within a Hamiltonian approach as a distinguished class of canonical transformations \cite{Gpap,GRG}. As a matter of fact, for this kind of approach, it is convenient to develop the theory in the interior of the BH, where the Hamiltonian evolution will truly take place in a timelike direction, with metric functions depending only on the parameter identified as the (interior) time coordinate, therefore simplifying the description. Consequently, the (Fourier) modes of the master functions will evolve dynamically in the interior \cite{Gpap,Master,GRG}. To relate this Hamiltonian dynamics to the exterior description, we can write the background dependence of the perturbed system solely in terms of background phase space variables and introduce a suitable symplectomorphism, i.e. a complex canonical transformation, such that it connects the interior background geometry to the exterior background geometry of the BH. As shown, e.g. in Refs. \cite{AOS,AOS2,AO,Leno}, it is remarkable that physical phase space functions such as the Hamiltonian remain real by means of their functions possessing an invariant range under this transformation or by passing from the positive real axis to the negative one. Thus, once in the exterior, where the roles of the timelike and radial coordinates are interchanged, we obtain a Hamiltonian interpreted as dictating the radial evolution of the system. In principle, the respective descriptions associated with different Fock representations in the BH interior can be extended to the exterior. 

In the present work, we will focus our attention on the quantum field theory obtained from the gauge invariant perturbations when the quantum corrections to the background are incorporated in an effective description for it. Classical backgrounds in General Relativity may be considered as a particular case in which no quantum correction is made. For this quantum field theory, a fundamental issue is the quantum equivalence of different representations when the infinite number of modes that compose the perturbations are taken into account. This includes the infinite possible positive integer values of the angular momentum label $l$ of the spherical harmonic defining the mode. Prior to examining isospectral features between different choices of master variables, it is therefore essential to determine whether the transformation relating the two sets of variables can be implemented as an operator that respects the quantum equivalence of the resulting descriptions. Does a classical DT admit a unitary implementation in the quantum representation chosen for the gauge invariant perturbations, or can the choice to perform the transformation before quantization affect the resulting quantum theory? If the answer to the first question is in the negative, should one first perform the classical canonical transformation and then quantize the resulting variables, or should one quantize the original system? The order would not affect the physics if a unitary transformation exists between the quantum descriptions of the two Darboux-related variables, establishing them to be equivalent. 

Let us emphasize that, while for a system with finitely many degrees of freedom, the unitary implementation of a given canonical transformation is generally much less problematic or trivial, the situation is more subtle when considering gravitational perturbations which comprise an infinite set, and in particular an infinite number of different angular momentum values. Even if the transformation is well-defined as a unitary map for each (fixed) individual value of $l$, its implementation across the infinitely many $l$-modes is not guaranteed to be unitary \cite{PRD83,DreamTeam}. Thinking about the classical DT as a map that changes the perturbative (master) variables (or as a Bogoliubov transformation, if preferred, between the associated creation and annihilation variables) and that, correspondingly, changes also the form of the perturbative Hamiltonian and equations of motion, it is natural to wonder about the existence of a unitary operator implementing the transformation in the quantum realm when all possible modes are taken into account. Addressing this issue is therefore a primary concern when determining whether the classical freedom to choose different master variables persists after quantization as a freedom to change to alternative equivalent representations.  

On the other hand, some previous works \cite{GnR,Beatles} argue that, in the present scenario of a nonrotating BH, a privileged family of equivalent quantum representations can be selected by respecting the background symmetries and by requiring unitarily implementable dynamics. By adopting one of these representations, the quantization can be claimed to be unique up to unitary equivalence. Moreover, the imposition of unitarily implementable dynamics in the form presented in Refs. \cite{GnR,Beatles} allows us to split the field dynamics into a part corresponding to the Heisenberg (unitary) dynamics associated with the field and another part that can be addressed as contributions arising from the evolution of the background itself\footnote{In this regard, the study of particle detectors provides an example of how these ideas can be applied, and in fact be beneficial, in curved spacetimes, as explored in Ref. \cite{ThePolice}.}.

In light of these considerations, the aim of this work is twofold. First, we demonstrate that canonical transformations associated with the DTs are unitarily implementable, even for effective backgrounds, so that the resulting quantum description is independent of whether the DT is performed prior to quantization or implemented at the quantum level. Our proof applies to a broad class of potentials that depend on the angular momentum label and exhibit asymptotic growth proportional to $l^2$ for large $l$. We further assume that, with respect to their local dependence, the potentials are continuous within the analyzed regions of the BH (and bounded in the boundaries of the exterior, so as not to affect the boundary conditions on quasinormal modes). We also impose a well-motivated asymptotic requirement (for large $l$) on a constant of motion appearing in the Riccati equation that determines the DT, and therefore the difference between potentials. Beyond these conditions, the proof is general and does not rely on any specific form of the potentials. 

In addition, we show that the privileged massless representation corresponding to master variables with unitarily implementable dynamics, while providing a representation also for the master variables on which the DTs act, is not unitarily related to the massless representation associated with this last set of variables, which is a different representation.
 
Finally, the second objective of this work arises from a natural question: can every representation in the unitary family selected by the criterion of unitary dynamics and symmetry invariance be reached through a canonical DT? We will show that the answer is negative.

The structure of this paper is organized as follows. In Sec. \ref{dtn}, we present the classical framework, i.e. we describe the nonrotating BH (interior) background, and concisely introduce the Hamiltonian formulation of geometric perturbations in terms of gauge invariant variables. We allow the background to evolve dynamically, in particular permitting quantum modifications within such dynamics that admit an effective Hamiltonian description. In that section, we also construct two canonical transformations that map variables adapted for unitary implementability of their dynamics into the master variables characteristic of the Darboux framework, alongside the canonical transformation directly associated with the Darboux map. Section \ref{UniD} briefly reviews the fundamental ideas and results about unitary implementability in Fock spaces. These ideas have guided prior work on unitarily equivalent representations, e.g. in the selection of the family of equivalent representations with unitarily implementable dynamics in the context of nonrotating BHs, and remain pivotal for our main objectives. In Sec. \ref{UnitImp}, we perform an asymptotic analysis to demonstrate that the DT admits a unitary quantum implementation. We further emphasize the role of variable selection in the context of the representation choice, showing that the transformation alone connecting the two considered sets of variables (those showing unitary dynamics and those acted upon by the DTs) fails to be unitarily implementable. In Sec. \ref{Unsaturated}, we provide an explicit counterexample illustrating that a DT cannot access all representations within the target family, leaving this freedom unsaturated. Finally, Sec. \ref{Conclusions} presents our concluding remarks. Throughout this work, we adopt Planck units so that $c = G = \hbar = 1$.

\section{Perturbed black holes and canonical Darboux transformations}\label{dtn}

As mentioned in the Introduction, although in principle what is strictly known as DTs are transformations traditionally formulated for master equations in the exterior geometry of the BH, they can, nevertheless, be extended to the interior region \cite{GRG}. We can formulate the classical perturbative theory in the interior and, upon encoding quantum corrections on the background in an effective manner, construct a quantum field theory for the gauge invariant perturbations by choosing an appropriate (family of equivalent) Fock representation(s). In this section, we introduce our classical setup, alongside the relevant canonical transformations: those associated with DTs \cite{Gpap,GRG}, as well as the transformations necessary to map the variables adapted for unitary implementability of the dynamics
into the master variables typical of the Darboux framework.

\subsection{The background model}

Let us introduce the spacetime describing the interior geometry of a nonrotating BH. The line element characterizing this background, foliated by homogeneous hypersurfaces in a manner isomorphic to Kantowski-Sachs cosmologies \cite{KS,Weber}, is of the form \cite{AOS,AOS2}
\begin{equation}\label{1.1}
        ds^2 = - \bN^2  \frac{p_b^2}{L_0^2} \, |p_{c}| \;d\chi^2 + \frac{p_b^2}{L_0^2 \;|p_c|} \; d\zeta^2 + |p_c|\; d\Omega^2_2,
\end{equation}
where $d\Omega_2^2 = d\theta^2 + \sin^2(\theta) d\varphi^2$ is the line element of the unit 2-sphere $S^2$, and $L_0$ is referred to as the compactification scale, which is a constant period of the $\zeta$-coordinate and is used as an infrared regulator, e.g. to avoid infinities in the integrations over $\zeta$. In addition, $\bN$ is the densitized lapse function (times $4\pi L_0$) over the volume element $\mathcal{V}=4\pi \,p_b\sqrt{p_c}$, and it is naturally interpreted as a Lagrange multiplier in the Hamiltonian formalism that describes evolution in the $\chi$-direction. In the interior of the BH, the $\chi$ and $\zeta$ coordinates are of timelike and spacelike nature, respectively, the former depending obviously on the choice of $\bN$. We have refrained from using the symbols $t$ and $r$ for the nonspherical coordinates in the interior, employing $\chi$ and $\zeta$ instead, in order to avoid confusion by the fact that the timelike coordinate in the interior, $\chi$, becomes spacelike in the exterior region of the BH (and conversely for $\zeta$). The metric functions $p_b(\chi)$ and $p_c(\chi)$ conforming to the line element above are variables directly related to densitized triad variables. They are canonically conjugate, respectively, to variables $b(\chi)$ and $c(\chi)$ (which can be identified with components of Ashtekar-Barbero connections \cite{AOS,AOS2,Master}).

These background variables satisfy the Poisson algebra 
\begin{equation}\label{Poisson}
    \{ b ,p_b \} = 1, \qquad \qquad \{c,p_c\}=2,
\end{equation}
with all the remaining Poisson brackets being identically zero. The dynamics (in the $\chi$-direction) of this background model is governed by the Hamiltonian\footnote{We treat the background geometry dynamically in the sense that its phase space variables are allowed to evolve via their corresponding Hamiltonian equations. If we neglect backreaction when perturbations are introduced, the background Hamiltonian generates such classical equations via Poisson brackets.}
\begin{equation}\label{theBack}
    \Hb [\bN] = -\bN \frac{L_0}{2} \left(\Omega_b^2 + 2\Omega_b \Omega_c + \frac{p_b^2}{L_0^2} \right),
\end{equation}
where we have defined $\Omega_a = (a \,p_a) /L_0$ for $a=b, \, c$. From the background Hamiltonian, it is straightforward to see that $\Omega_c$ is a constant of motion (at least if backreaction is ignored), interpretable as the BH mass. 

\subsection{Geometry perturbations}

Given the topology of the constant-$\chi$ sections of our model, it is natural to decompose any field perturbation into eigenmodes of the Laplace-Beltrami operator for $S^1\times S^2$, with $2\pi \zeta/L_0\in S^1$.  A complete basis of eigenmodes is given by the product of Fourier modes for $S^1$, which are labeled by the Fourier frequency $\omega_n=2\pi |n|/L_0$ with $n$ taking integer values, and spherical harmonics for $S^2$, labeled by their angular momentum $l$ and the azimuthal number $m$. The angular momentum $l$ ranges over all nonnegative integer values, and $m$ can take integer values from $-l$ to $l$. The scalar-vector-tensor decomposition in spherical harmonics of the $3+1$ form of the metric results, respectively, in perturbations of the lapse function, the shift vector, and the three-metric (including its conjugate momentum) induced on sections along the foliation in the $\chi$-direction \cite{Axial,Gpap,GRG}.

The perturbations can be described by the $\chi$-dependent coefficients of their expansion in the aforementioned basis of modes, and will thus have associated labels $\n=(n,l,m)$. To avoid dealing with complex relations, we take the complete basis to be real, but certainly the ultimate result is independent of this choice. Given the compact topology of the considered sections, we can isolate the zero modes and treat them separately. Hence, in the rest of this article, zero modes will not be included in sums over the perturbative mode labels. These modes will be treated exactly in our description at the order of our perturbative truncation, and in particular independently of the perturbations if backreaction is directly ignored. Moreover, since modes with $l$ strictly less than two are not radiative gravitational degrees of freedom, we remove these modes from our considerations in the following, focusing our attention on the infinite set of modes with $l\geq 2$. 

For our perturbative analysis, we truncate the action at quadratic order in the perturbations \cite{Axial,Gpap,Master}. The truncated action in Hamiltonian form is composed of a Legendre term, which specifies the canonical structure and Poisson brackets of our background and perturbative variables, and a constraint part. The constraints split into two sets. On the one hand, we get perturbative constraints, arising from the linear perturbation of the diffeomorphism constraints. They generate perturbative gauge transformations, and possess Lagrange multipliers that are the mode coefficients of the perturbation of the lapse function and the shift vector \cite{Gpap,Master,GRG}. On the other hand, given the symmetries of our background, we find only a global constraint. It is the zero mode contribution to the Hamiltonian gravitational constraint. This is a sum of two terms: the background Hamiltonian and a perturbative Hamiltonian that is quadratic in the perturbations (therefore, when backreaction is neglected, the background Hamiltonian appears as a constraint exclusively for the background model). The (zero mode of) the lapse function, $N$, is the Lagrange multiplier of this global constraint. 

The perturbative contributions to the constraints in the truncated action can be classified according to their behavior under parity transformation on the sphere, $(\theta,\phi)\mapsto(\pi-\theta,\phi+\pi)$. There are two types of terms: the axial and polar contributions \cite{Frolov}. On the other hand, it is possible to introduce a background-dependent canonical (linear) transformation\footnote{This canonical transformation of the perturbative variables, being background dependent, spoils the commutation under Poisson brackets of the perturbations with the background. However, this can be solved by suitably modifying the background zero modes with quadratic corrections of the perturbations, restoring the global canonical structure at our perturbative truncation order in the action \cite{Gpap,Master}.} of the perturbative variables to separate them into perturbative gauge degrees of freedom and physical, gauge invariant perturbations \cite{Gpap,Master,GRG}. The latter are characterized by commuting under Poisson brackets with the perturbative constraints. A consequence of the background dependence of this change of perturbative variables is that the perturbative Hamiltonian is modified, because part of the original perturbative dynamics is redistributed to the background.   

Even though the above procedure allows to construct a complete canonical set of gauge invariants for the perturbations, we can go a step further and introduce one more background-dependent canonical transformation of these gauge invariants to make them coincide with the modes of standard master functions employed in BH perturbation theory (and their canonical momenta). In consonance with our previous remarks, this change also allows us to bring the perturbative (gauge invariant) Hamiltonian into a more convenient form, diagonal in configuration and momentum variables and with a fixed (possibly background-dependent) coefficient for the squared momentum terms \cite{Gpap,Master,GRG}. For instance, we can introduce the so-called Gerlach-Sengupta (GS) master variables \cite{GS1,GS2}. 

For the purpose of the present article, it is sufficient to consider the axial sector of the perturbations as the center of our discussion, given its simplicity. The polar sector can be formally related to it through a specific DT (the Chandrasekhar transformation \cite{Chandra3}), precisely the class of canonical transformations that we want to study in this work. The resulting axial GS Hamiltonian $\mathrm{H}_{\GS}$ will serve as the pivotal Hamiltonian from which we start the presentation of our analysis. In such analysis, we choose the notation $\tau$ for the parameter associated with the dynamical evolution when the lapse function $N$ is set equal to unity. For the sake of brevity, we refer to this parameter as the unit-lapse time. Then, the perturbative Hamiltonian takes the following form \cite{Gpap}:
\begin{equation}\label{sax}
    \mathrm{H}_{\GS} = \frac{1}{2\kappa}\sum_{\n} \left(\left[\Pi_{\n}^{\GS} \right]^2+\left(\omega_n^2p_c^2 + l(l+1) \frac{p_b^2}{L_0^2} +\mathscr{s}^{\p} \right) \left[\phi_{\n}^{\GS} \right]^2\right).
\end{equation}
Here, we have adopted the notation $\kappa=16\pi$, the superindex $\GS$ refers of course to Gerlach-Sengupta, $\Pi^{\GS}_{\n}$ is a canonical momentum of the configuration variable $\phi_{\n}^{\GS}$, with the convention that the nontrivial canonical Poisson brackets for the perturbations are given as $\{\phi_{\n}^{\GS}, \Pi^{\GS}_{\breve{\n}}\}=\kappa \delta_{\n\breve{\n}}$, and $\mathscr{s}^{\p}$ is a background-dependent but $\n$-independent contribution to the potential. In terms of background phase space variables and their dynamical derivatives, it adopts the expression
\begin{equation}\label{Schwar}
    \begin{aligned}
\mathscr{s}^{\p} &= 6 \Omega_b \Omega_c + \frac{1}{2}\frac{|p_c|^{\prime\prime}}{|p_c|} - \frac{3}{4}\frac{(|p_c|^{\prime})^2}{|p_c|^2}, 
\end{aligned}
\end{equation}
where the prime denotes the derivative with respect to the time $\tau$. We recall that, in classical General Relativity and at the considered order in our perturbative description, this derivative is given by the Poisson bracket with the background Hamiltonian with unit lapse. On the other hand, for effective backgrounds admitting a Hamiltonian formalism, the background Hamiltonian generating this derivative must be replaced with an effective Hamiltonian, but no other substantial differences arise. This fact will ensure that the results of our study remain fully applicable for such effective backgrounds. Similar considerations apply to other derivatives of the background variables that will appear later in our discussion. Note also that, both in the classical and in effective cases, the resulting Poisson brackets of background phase space functions provide again functions on this phase space. Finally, let us comment that the two last summands in Eq. \eqref{Schwar} are one half the Schwarzian derivative \cite{Sch} of the primitive of $|p_c|$. 

Evaluating the derivatives in classical General Relativity, for example, for which $|p_c|^{\prime}= 2\Omega_b |p_c|$, we obtain
\begin{equation}
    \begin{aligned}
        \mathscr{s}^{\p}&= 6\Omega_b\Omega_c-\Omega_b^2-\frac{p_b^2}{L_0^2}.
    \end{aligned}
\end{equation}
Moreover, on the Schwarzschild solution to General Relativity, the lapse function and background variables take the on-shell form \cite{Gpap,Master,GRG}
\begin{equation}
   \bN^{-1}= \frac{p_b^2}{L_0^2}=-\chi^2\left(1-\frac{2M}{\chi}\right), \quad |p_c|=\chi^2, \quad \Omega_b=\chi\left(1-\frac{2M}{\chi}\right), \quad \Omega_c=M,
\end{equation} 
where $M$ denotes the BH mass. In particular, with the change of notation $\chi=r$ in the exterior region, substituting these expressions into Eq. \eqref{1.1} reproduces the familiar Schwarzschild line element. Evaluating $\mathscr{s}^{\p}$ on this solution gives
\begin{equation}
    \mathscr{s}^{\p}=8M\chi\left( 1- \frac{2M}{\chi}\right).
\end{equation}

\subsection{Canonical transformation leading to unitary dynamics}

In order to inspect the unitary implementability of the DTs, let us now construct a canonical transformation $\left(\phi_{\n}^{\GS},\Pi_{\n}^{\GS}\right)\xrightarrow{\mathscr{R}_{\l}}(\tilde{\phi}_{\n}^{\GS},\tilde{\Pi}_{\n}^{\GS})$, leading to new variables $(\tilde{\phi}_{\n}^{\GS},\tilde{\Pi}_{\n}^{\GS})$, with respect to which the perturbative Hamiltonian takes the form of an infinite sum of harmonic oscillators with background-dependent masses.

Let us start with some notation that will be useful in our study. We define a \textit{wavenumber} label and a normalized \textit{$l$-component} label $\l$ as
\begin{equation}\label{kl}
    k^2 = \omega_n^2+l(l+1), \qquad \qquad \l= \frac{\sqrt{l(l+1)}}{k},
\end{equation}
where we have left the zero mode apart as mentioned earlier. We note that $0<\l^2= 1- (\omega_n^2/k^2)\leq 1$, (with $l\geq 2$). Using these definitions, we introduce the following quantity:
\begin{equation}
    \b^2 = \left(1-\l^2\right) p_c^2 + \l^2 \frac{p_b^2}{L_0^2}.
\end{equation}
Since $\l \in (0,1]$, the phase space function $\b$ remains bounded from above and below at any given time when the pair $(n,l)$ runs over all possible values \cite{GnR}, provided that (here and in the rest of our analysis) we assume the background phase space variables to be continuous in $\tau$ within the studied region. 

The aforementioned canonical transformation is then given by
\begin{equation}
    \begin{pmatrix}
        \tilde{\phi}_{\n}^{\GS}\\
        \tilde{\Pi}_{\n}^{\GS}
    \end{pmatrix}
    =\mathscr{R}_{\l}
    \begin{pmatrix}
        \phi_{\n}^{\GS}\\
        \Pi_{\n}^{\GS}
    \end{pmatrix}, \quad \text{with} \quad 
    \mathscr{R}_{\l}=
    \begin{pmatrix}
        \sqrt{\b} & 0 \\
        \frac{1}{2} \frac{\b^{\prime}}{\b^{3/2}}    & \frac{1}{\sqrt{\b}}
    \end{pmatrix},
\end{equation}
where again the prime denotes the derivative with respect to the unit-lapse time $\tau$. The resulting Hamiltonian is 
\begin{equation}\label{tildeHam}
    \tilde{\mathrm{H}}_{\GS} = \frac{1}{2\kappa} \sum_{\n} \b \left(\left[ \tilde{\Pi}_{\n}^{\GS} \right]^2 + \left(k^2 + s_{\l}^{\p}\right)  \left[ \tilde{\phi}_{\n}^{\GS} \right]^2\right),
\end{equation}
where
\begin{equation}
\begin{aligned}\label{smass}
    s_{\l}^{\mathrm{ax}} &= \s+ \frac{\mathscr{s}^{\p}}{\b^{2}}, 
    \end{aligned}
\end{equation}
and $\s$ corresponds to the mass term of the modes of a free Klein-Gordon field in a Kantowski-Sachs geometry, which is given by $-1/(2\b^2)$ times the Schwarzian derivative of the primitive of $\b$, namely
\begin{equation}
    \s = \frac{3}{4}\frac{(\b^{\prime})^2}{\b^4}-\frac{1}{2}\frac{\b^{\prime\prime}}{\b^3}.\end{equation}
Alternatively, we can express it in terms of background phase space variables using the Hamiltonian equations for the background (at our perturbative truncation order). For instance, in classical General Relativity,
\begin{equation}
    \s =  \frac{1}{\b^2} \left( \Omega_b^2+\frac{p_b^2}{L_0^2} \right)- \frac{\l^2 p_b^2}{2\b^4L_0^{2}} \left[  4\Omega_b^2-8\Omega_c^2-12   \Omega_b \Omega_c+\frac{p_b^2}{L_0^2}\right] + \frac{5}{4}\frac{\l^4 p_b^4}{\b^6L_0^{4}}\left(\Omega_b-2\Omega_c\right)^2.
\end{equation}

Given the resemblance of the resulting Hamiltonian to the one of an infinite tower of generalized oscillators, up to the global factor $\b/\kappa$, we will refer to these variables as the \textit{oscillator-adapted} ($\GS$) variables. Along the same line of discussion, it is important to note that Eq. \eqref{tildeHam} presents practically the same form and characteristics as the Hamiltonian studied in Ref. \cite{GnR,Beatles}, namely, besides the aforementioned global $\b$-factor, the only phase space function encoding the gravitational information is the mass function $s_{\l}$, which, as in that case, is found to be bounded from above and below with respect to the mode labels $\n$ at each of the analyzed times (under the assumed behavior of the background variables), and depends purely on background functions and on the (normalized) label $\l$. Hence, although the aforementioned reference is, in principle, an analysis considering a scalar field that acts as a perturbation, it encompasses broader field perturbations as the present ones. Therefore, the results presented in Refs. \cite{GnR,Beatles} apply to the geometry perturbations considered here. Concretely, they imply that a judicious choice of Fock representation for the field description of our system in the (effective) interior of a nonrotating BH is achieved such that the Heisenberg dynamics of the corresponding creation and annihilation operators is unitarily implemented at the quantum level. Invariance of the representation under the background isometries is assumed. Such a representation of the canonical commutation relations is, moreover, unique up to unitary equivalence. Obviously, this Fock quantization leads then to a(n essentially) unique choice of classical variables of creation and annihilation type.

\subsection{Cunningham-Price-Moncrief variables and Darboux canonical transformations}

The mode variables corresponding to the CPM master function alongside canonical momenta for them are given in Ref. \cite{Gpap}. They are obtained from the Gerlach-Sengupta variables by means of a background-dependent canonical transformation. Let us denote these CPM variables as $(\phi_{\n}^{\M},\Pi_{\n}^{\M})$. The aforementioned transformation, $\left(\phi_{\n}^{\GS},\Pi_{\n}^{\GS}\right)\xrightarrow{\mathscr{S}}\left(\phi_{\n}^{\M},\Pi_{\n}^{\M}\right)$, adopts the expression
\begin{equation}
    \begin{pmatrix}
        \phi_{\n}^{\M} \\
\Pi_{\n}^{\M} 
    \end{pmatrix}= \mathscr{S} 
    \begin{pmatrix}
        \phi_{\n}^{\GS} \\
\Pi_{\n}^{\GS}
    \end{pmatrix}, \quad \text{with} \quad \mathscr{S}= \frac{1}{\sqrt{|p_c|}}
    \begin{pmatrix}
        |p_c| & 0 \\
         \frac{|p_c|^{\prime}}{2|p_c|} & 1
    \end{pmatrix}.
\end{equation}
In the particular case of classical General Relativity, we recall that $ |p_c|^{\prime}/(2|p_c|)=\Omega_b$. The Hamiltonian (in the unit-lapse time) for these CPM variables is
\begin{equation}\label{CPMZ}
    \mathrm{H}_{\M} =\frac{|p_c|}{2\kappa} \sum_{\n}  \left( \left[\Pi_{\n}^{\M}\right]^2 +\left(\omega_n^2 - V_{l}^{\p}\right)\left[\phi_{\n}^{\M}\right]^2\right), 
\end{equation}
where $V_{l}^{\p}$ is the Regge-Wheeler potential expressed in terms of the background phase space variables \cite{Gpap,GRG}, namely
\begin{equation}\label{potentials}
    \begin{aligned}
        V^{\p}_{l} &= -\frac{1}{p_c^2} \left[l(l+1) \frac{p_b^2}{L_0^2} + 6\Omega_b \Omega_c \right].
    \end{aligned}
\end{equation}
Evaluating this potential, for instance, on the Schwarzschild solution of classical General Relativity leads to the familiar expression
\begin{equation}
V_{l}^{\mathrm{RW}}=\left( 1- \frac{2M}{\chi}\right) \left[ \frac{l(l+1)}{\chi^2} -\frac{6M}{\chi^3} \right].
\end{equation}

On the other hand, DTs, as employed, e.g. in General Relativity, are characterized by relating the Regge-Wheeler master equation satisfied by the CPM master function to equations of similar form but with distinct potentials. A particular example is the Zerilli equation, which is precisely the equation satisfied by the Zerilli-Moncrief master function that characterizes the gauge invariant polar perturbations \cite{Ze,Mo} (in this case, the map is provided by the Chandrasekhar transformation \cite{Chandra3}). More concretely, let $d\tau_{\star}=|p_c|d\tau$ be the \textit{tortoise} coordinate\footnote{This coordinate is timelike in the interior of the BH. By matching Eq. \eqref{1.1} with the Schwarzschild interior metric, we can derive the relation between $|p_c|$ and the Schwarzschild interior timelike coordinate. The definition of the tortoise coordinate $\tau_{\star}$ and its relation to the standard tortoise coordinate of the exterior becomes then evident, as discussed e.g. in Ref. \cite{BlackPumas}.}. For a differential operator $L_{nl}=(\partial_{\tau_\star}^2 + \omega_n^2+V_{l}^{\p})$ acting on a mode, say $Q_{\n}$ (where the specific value of the label $m$ is completely irrelevant), so that $L_{nl}Q_{\n}=0$, we ask whether, for given angular momentum $l$, we can transform every solution $Q_{\n}$ into a solution $\tilde{Q}_{\n}=A_{l}Q_{\n}$ that satisfies a similar equation $\tilde{L}_{nl}\tilde{Q}_{\n}=0$, where $\tilde{L}_{nl}=(\partial_{\tau_{\star}}^2 + \omega_n^2+v_{l})$. This transformation is defined via the Darboux intertwining relation\footnote{The intertwining relation results in the usual operator transformation $\tilde{L}_{nl}=A_{l}L_{nl}A^{-1}_{l}$ when the inverse $A_{l}^{-1}$ exists.}: $\tilde{L}_{nl}A_{l}=A_{l}L_{nl}$, and $v_l$ is a new potential. If we let $A_{l}$ be simply a ($\tau_{\star}$-dependent) rescaling, the transformation becomes a trivial change of variables in the resulting equation. The simplest nontrivial local (in the tortoise coordinate) transformation is a first-order differential operator $A_{l}= \partial_{\tau_{\star}}+g_{l}(\tau_{\star})$. By expanding both sides of the Darboux intertwining relation and matching terms, the resulting constraint is found to be that the scaling\footnote{This scaling is assumed not to diverge at the boundaries of the exterior region.} $g_l$ must satisfy the Riccati equation $\partial_{\tau_{\star}}\,g_l-g_l^2+V_{l}^{\p}+c_{l}=0$, which characterizes each DT. Here, $c_{l}$ is a constant of motion. Finally, the relation between the two considered potentials is then $v_{l}=V_{l}^{\p}+2\partial_{\tau_{\star}}\,g_{l}$ \cite{Intertwining1, Intertwining2}.

Furthermore, it was proved in Refs. \cite{Gpap,Master,GRG} that there exists a one-to-one correspondence between DTs and canonical transformations that preserve the diagonal harmonic-oscillator form of the Hamiltonian in the tortoise coordinate presented in Eq. \eqref{CPMZ}. The family of canonical transformations that characterize DTs within a Hamiltonian framework are transformations $(\phi_{\n}^{\M},\Pi_{\n}^{\M})\xrightarrow{\mathscr{D}_{nl}}(\phi_{\n}^{\MM},\Pi_{\n}^{\MM})$ of the form
\begin{equation}\label{Darboux}
    \begin{pmatrix}
       \phi_{\n}^{\MM}\\
       \Pi_{\n}^{\MM}
    \end{pmatrix} = \mathscr{D}_{nl}
    \begin{pmatrix}
        \phi_{\n}^{\M}\\
       \Pi_{\n}^{\M}
    \end{pmatrix} , \quad \text{with} \quad \mathscr{D}_{nl}= -\frac{1}{(\omega_n^2+c_l)^{1/2}} \begin{pmatrix}
        g_{l} & \;\;\; 1 \\
        g_{l}^2-\omega_{n}^2-c_{l} & \;\;\; g_{l}
    \end{pmatrix},
\end{equation}
where we emphasize that the constant $c_l$ is independent of the Fourier frequency. The connection with the DT introduced above can be seen directly from the first row of this canonical transformation. Using the Hamiltonian equations of motion, the momentum can be replaced by the time derivative of the configuration variable, so that the transformed configuration variable is proportional to $\partial_{\tau_{\star}}\phi_{\n}^{\M}+g_l\,\phi_{\n}^{\M}$. Thus, the canonical transformation acts on the configuration variable precisely as the first-order Darboux operator introduced above. In other words, the canonical transformation maps the original master variable to the Darboux-transformed master variable, while at the same time transforming its conjugate momentum so that the transformation remains canonical.

The setting is therefore ready to display how DTs act in terms of the privileged set of oscillator-adapted variables introduced in the previous subsection. Given the relation between the different sets of variables, we  obtain a linear canonical transformation of the form
 $(\tilde{\phi}_{\n}^{\GS},\tilde{\Pi}_{\n}^{\GS})\xrightarrow{\mathscr{B}_{nl}}(\tilde{\phi}_{\n}^{\GGSS},\tilde{\Pi}_{\n}^{\GGSS})$, with
\begin{equation}
    \mathscr{B}_{nl} = \left(\mathscr{R}_{\l}\,\mathscr{S}^{-1}\right)\mathscr{D}_{nl}\left(\mathscr{R}_{\l}\,\mathscr{S}^{-1}\right)^{-1}.
\end{equation}
Explicitly, this acts as
\begin{equation} \label{Bogoliu}
    \begin{pmatrix}
        \tilde{\phi}_{\n}^{\GGSS}\\
        \tilde{\Pi}_{\n}^{\GGSS}
    \end{pmatrix} = \mathscr{B}_{nl}
    \begin{pmatrix}
         \tilde{\phi}_{\n}^{\GS}\\
        \tilde{\Pi}_{\n}^{\GS}     
    \end{pmatrix} \quad \text{with} \quad \mathscr{B}_{nl} = -\frac{1}{(\omega_n^2+c_l)^{1/2}} \begin{pmatrix}
        g_{l}-\mathcal{G}_{\l}\,\bs^{1/2} 
        & \;\; \bs \\
       \mathcal{D}_{nl} & \;\; g_{l}+\mathcal{G}_{\l}\,\bs^{1/2}
    \end{pmatrix},
\end{equation}
where the $\star$-symbol over $\b$ stands for $\bs=\b/|p_c|$. Taking the derivative of this definition yields $\b^{\prime} = |p_c|^{\prime}\,\bs + |p_c|\,\bs^{\prime}$. In particular, in classical General Relativity, where the Hamiltonian $\Hb [\bN=1]$ dictates the background evolution, we obtain the identity $\bs^{\prime}\, |p_c| = \b^{\prime} - 2\b\Omega_b$. Moreover, in the transformation above, we have defined
\begin{equation}
\mathcal{G}_{\l} = \frac{\bs^{\prime}}{2\,|p_c|\,\bs^{3/2}},\qquad \qquad
    \mathcal{D}_{nl}=\frac{\left(g_{l}^2-\omega_n^2-c_{l}\right)}{\bs} - \mathcal{G}_{\l}^2.
\end{equation}

\section{Unitarity Conditions}\label{UniD}

Now that we have established how DTs act on the variables $(\tilde{\phi}_{\n}^{\GS},\tilde{\Pi}_{\n}^{\GS})$, we can check under what conditions they are unitarily implementable if we consider the simultaneous transformation of all modes, including all possible values of the angular momentum. For this discussion, we adopt a privileged Fock representation selected (up to unitary equivalence) by invariance under classical symmetries and the existence of a nontrivial quantum unitary Heisenberg evolution \cite{GnR,Beatles}. Concretely, we follow the arguments and results of Ref. \cite{GnR}. 

First, let us choose such a specific Fock representation within the aforementioned unique family with unitary dynamics. As it is well-known, a Fock representation for a linear field theory is determined by a so-called complex structure in the classical phase space, i.e. a symplectic transformation 
$J$ (i.e. a transformation compatible with the symplectic structure) such that its square is minus the identity map, $J^2=-\mathbf{1}$ \cite{Wald,Wald2,Winitzki}. Such a linear transformation is fully determined by its set of (complex) eigenvectors, corresponding in this case to (infinitely degenerated) eigenvalues $\pm i$, as follows from the condition $J^2=-\mathbf{1}$. Physically, these eigenvectors correspond, upon quantization, to creation and annihilation operators, and it follows that choosing a particular Fock representation amounts to declaring which set of classical (complex) variables are to be promoted to creation and annihilation operators in Fock space \cite{Wald,AshtekarMagnon}.
  
The family of representations selected by unitary Heisenberg dynamics, attainable for the mode variables $(\tilde{\phi}_{\n}^{\GS},\tilde{\Pi}_{\n}^{\GS})$, is known to contain the massless representation corresponding to these canonical pairs, as was proved in Ref. \cite{GnR}. Given its simplicity, let us then choose this massless Fock representation for our considerations. Unitary equivalence ensures that results about unitary implementability are independent of this choice. Such a massless representation is determined by the following creation and annihilation variables \cite{GnR,Beatles}:
\begin{equation}\label{reprensation}
	\begin{pmatrix}
		a_{\n} \\ a_{\n}^{*}
	\end{pmatrix} = 
	\tilde{\mathscr{F}}_{nl}^{\mathrm{o}}
	\begin{pmatrix}
		\tilde{\phi}_{\n}^{\GS} \\ \tilde{\Pi}_{\n}^{\GS}
	\end{pmatrix}, \qquad \text{where} \quad \tilde{\mathscr{F}}_{nl}^{\mathrm{o}} = 
	\begin{pmatrix}
	\tilde{f}_{nl}     & \tilde{g}_{nl}  \\
		\tilde{f}_{nl}^{*}  & \tilde{g}_{nl}^{*} 
	\end{pmatrix} ,
\end{equation}

\begin{equation} \label{z11}
\text{and} \quad \quad	\tilde{f}_{nl} = \sqrt{\frac{k}{2}}=\frac{1}{\sqrt{2}} \left[\omega_n^2+l(l+1)\right]^{1/4}, \qquad \tilde{g}_{nl} = \frac{i}{\sqrt{2k}}=\frac{i}{\sqrt{2}\left[\omega_n^2+l(l+1)\right]^{1/4}} ,
\end{equation}
where the asterisk as a superscript denotes the complex conjugate. 

Unitary implementation of a canonical transformation implies that there exists a unitary operator providing a quantum version of the transformation in question. In the present case, we consider the implementation of the set of canonical transformations corresponding to the DTs given by
$\mathscr{B}_{nl}$ in Eq. \eqref{Bogoliu} above. By introducing our creation and annihilation variables, our Darboux-Bogoliubov transformation, in terms of the creation and annihilation variables for the chosen massless representation, takes the form 
\begin{equation} \label{Bogo}
	\begin{aligned}
		\begin{pmatrix}
			\tilde{a}_{\n} \\ \tilde{a}_{\n}^{*}
		\end{pmatrix} = \mathscr{A}_{nl} 
		\begin{pmatrix}
			a_{\n} \\ a_{\n}^{*}
		\end{pmatrix}~, 
	\end{aligned}
\end{equation}
where
\begin{equation}\label{BogoD}
	\begin{aligned}
		\mathscr{A}_{nl} = 
		\begin{pmatrix}
			\alpha_{nl}    & \beta_{nl} \\
			\beta_{nl}^{*} & \alpha_{nl}^{*}  
		\end{pmatrix}
		= \tilde{\mathscr{F}}_{nl}^{\mathrm{o}}\;\mathscr{B}_{nl}\left(\tilde{\mathscr{F}}_{nl}^{\mathrm{o}}\right)^{-1}.
	\end{aligned}
\end{equation}
Let $\hat{a}_{\n}^{\dagger}$ and  $\hat{a}_{\n}$ denote the creation and annihilation operators in Fock space, i.e. the quantum counterparts of the classical variables $a_{\n}^*$ and $a_{\n}$. Unitary implementation of the DTs then means that there exist unitary operators $\hat U$ such that, for all $\n$,
\begin{equation} \label{unitaryop}
	\hat U \hat a_{\n} \hat U^{-1}= \alpha_{nl} \hat a_{\n} + 
	\beta_{nl} \hat a_{\n}^{\dagger},
\end{equation}
and analogously for $\hat a_{\n}^{\dagger}$. Whereas the right-hand side of Eq.\ \eqref{unitaryop} is always well-defined, the existence of $\hat U$ is not ensured, in principle.  In fact, its existence is tantamount to the unitary equivalence of the two representations defined by the two sets of creation and annihilation pairs, namely the original one \eqref{reprensation} and the transformed one, by means of the Bogoliubov transformation \eqref{Bogo}.

General results \cite{Shale,HR} require as a necessary and sufficient condition for unitary implementation that the $\beta$-coefficients of the studied transformation have square summable complex norms. Mathematically, this condition amounts to demanding that the antilinear part of the transformation be Hilbert-Schmidt \cite{PRD83}. 

Therefore, in the current case, we need
\begin{equation}\label{sum} 
	\sum_{\n}  |\beta_{nl}|^2 = \sum_{nl} (2l+1) |\beta_{nl}|^2< \infty ,
\end{equation}
 where the factor $2l+1$ accounts for the degeneracy in the label $m$. This condition can be thought of physically as requiring a finite number of particles when comparing the vacua of the two representations \cite{Wald,Winitzki,Beatles}. 

\section{Unitary implementability}\label{UnitImp}

We will now show that DTs are indeed unitarily implementable, under typical conditions. To discuss the unitary implementability of DTs, let us then consider the Hilbert-Schmidt criterion of Eq. \eqref{sum}. Explicitly, the $\beta_{nl}$ coefficient of Eq. \eqref{BogoD} yields 
\begin{equation}\label{beta}
\begin{aligned}
    \beta_{nl}&=
     \frac{1}{\sqrt{ \omega_n^2+c_{l}}} \left( \mathcal{G}_{\l}\,\bs^{1/2} - \frac{i}{2} \left[ k\bs + \frac{\mathcal{D}_{nl}}{k} \right]\right)\\
     &= \frac{1}{\sqrt{ \omega_n^2+c_{l}}} \left( \mathcal{G}_{\l}\,\bs^{1/2} - \frac{i}{2k\bs} \left[k^2 \bs^2 +  V_{l}^{\p}+\partial_{\tau_{\star}}g_{l}-\omega_{n}^2 -  \mathcal{G}_{l}^2\,\bs \right]\right),
\end{aligned}
\end{equation}
where we have substituted the definition of $\mathcal{D}_{nl}$, and used the Riccati equation $\partial_{\tau{\star}} \, g_l-g_l^2+V_{l}^{\p}+c_{l}=0$ to obtain the last equality. Since all (phase space) functions in the above expression are supposed to be well-behaved in the interval of the evolution parameter under consideration, the summability of these coefficients in square norm reduces to analyzing their behavior in the ultraviolet sector, i.e. the region in the space of labels where $k$ grows unboundedly large, allowing that  $n$ and/or $l$ can tend to infinity.

\subsection{Asymptotic Analysis}\label{Asymp}

In order to study the asymptotic behavior of expression \eqref{beta}, let us consider the different terms in its numerator. Noting that $k^2 \bs^2 = \omega_n^2 + l(l+1) p_b^2/(L_0^2\,p_c^2)$ we can simplify the $\omega_n^2$-terms. In addition, $V_{l}^{\p}$ behaves asymptotically as $-p_b^2 l^2/(L_0^2\,p_c^2)$ for large $l$. We will assume that the term $V_{l}^{\p} + \partial_{\tau_{\star}}\,g_{l}$ has also this behavior, i.e. $V_{l}^{\p}$ is dominant over $\partial_{\tau_ {\star}} g_{l}$. In fact, this assumption can be further refined and substantiated. Consider the Riccati equation that $g_{l}$ satisfies, namely $\partial_{\tau_{\star}}\,g_{l}-g_{l}^2+V_{l}^{\p} +c_{l}=0$, and recall that $V_{l}^{\p}/l^2$ remains bounded at each analyzed instant of (interior) time when $l$ runs over the integers. Then, the constant of motion $c_{l}$ can dominate for large $l$ in the Riccati equation against $\partial_{\tau_{\star}}\,g_{l}$ and $V_{l}^{\p}$ if it is of sufficiently high asymptotic order in $l$. Solving the equation order by order we obtain that $g_{l}=\sqrt{c_{l}}+g_{l,\mathrm{sub}}$, where $g_{l,\mathrm{sub}}$ contains subdominant contributions for large $l$. In particular, note then that $\partial_{\tau_{\star}}g_{l}=\partial_{\tau_{\star}}g_{l,\mathrm{sub}}$ is indeed subdominant, because $c_{l}$ is a constant of motion. Balancing the next order, we obtain $g_{l,\mathrm{sub}}=V_{l}^{\p}/\sqrt{4c_{l}}$ up to subleading terms. As previously mentioned, the new potential is defined as $v_l=V_l^{\p}+2\partial_{\tau_{\star}}\,g_l$. Since $\partial_{\tau_{\star}}g_l = \partial_{\tau_{\star}}V_{l}^{\p}/\sqrt{4c_{l}}$ at leading asymptotic order in $l$, we notice that the correction to the potential is of the same order as $g_{l,\mathrm{sub}}$ (assuming that the $\tau_{\star}$-derivative commutes with the asymptotic series for $g_{l}$). Hence, $V_{l}^{\p}$ dominates over $\partial_{\tau_\star}\,g_{l}$, and the new potential $v_l$ is in fact of the same asymptotic order in $l$ as the Regge-Wheeler potential $V_{l}^{\p}$. We now further concretize our assumption about the asymptotic behavior of $c_{l}$ for large $l$, accepting that it grows faster than $l^4$. It then follows that $g_{l,\mathrm{sub}}$ tends to zero in that asymptotic limit, and as a consequence the same applies to the difference of potentials. Transformations satisfying this condition that $c_{l}$ be of asymptotic order dominant over $l^4$ include, in particular, the Chandrasekhar transformation. For General Relativity, the expression for $c_{l}$ can be found in Ref. \cite{GRG} (explicitly, we have $c_{l} =[p_b^4(l+2)!/(12L_0^4\Omega_b^2\Omega_c |p_c| (l-2)!)]^2$) and behaves asymptotically as $l^8$. Moreover, a careful analysis in order to extend $c_{l}$ to effective backgrounds, such as those proposed in Refs. \cite{AOS,AOS2} to include quantum modifications, confirms that this asymptotic order is not altered by such corrections.

On the other hand, since the phase space function $\b$ is bounded from above and below with respect to the labels $n$ and $l$, this boundedness extends to $\bs$, and consequently to $\mathcal{G}_{\l}$. Thus, the first term in the parentheses in Eq. \eqref{beta} is bounded, and the last term is subdominant with respect to it in the ultraviolet sector. Taking all the above into account, the summability in square norm of the $\beta$-coefficients \eqref{beta} amounts to that of the following terms: 
\begin{equation}\label{2Sums}
 \frac{1}{\sqrt{ \omega_n^2+c_{l}}}\left| \mathcal{G}_{\l}\, \bs^{1/2} - i\frac{p_b^2l^2}{2L_0^2\,p_c^2\,\bs k} \right| \leq \left|\frac{L_0 \mathcal{G}_{\l}\,\bs^{1/2}}{\sqrt{4\pi^2n^2 +L_0^2c_{l}}} \right| + \left| \frac{p_b^2l^2}{2p_c^2\bs \sqrt{4\pi^2 n^2+L_0^2c_{l}}\sqrt{4\pi^2 n^2+ L_0^2 l(l+1)}\,}\right|.
\end{equation}
In the last triangular inequality, we have also substituted the definition of $\omega_n$ and $k$ in terms of $n$ and $l$.

Let us focus first on the square summability of the first term. Summing over the label $m$ produces a factor of $2l+1$. But this factor is precisely the number of integers between a value of the square angular momentum label $l^2$ and the next valid value of this square, $(l+1)^2$. Let us then call $y=l^2$, and rewrite the sum over all modes of the considered term as
\begin{equation}\label{FirstSum}
\sum_{nl}\frac{(2l+1) L_0^2 (\mathcal{G}_{\l})^2\,\bs}{4\pi^2n^2 +L_0^2c_{l}} = \sum_n \sum_y  \frac{L_0^2 (\mathcal{G}_{\l})^2\,\bs}{4\pi^2n^2 +L_0^2 c_{[l]_{\text{ent}}}}, 
\end{equation}
where $[l]_{\text{ent}}$ is the integer part of $l=\sqrt{y}$ and we are summing in $y$ over all integers (greater than 4). Note that the numerator of the last expression is bounded when the labels $(n,y)$ run over all possible values at every instant of the (interior) time coordinate. In consonance with the asymptotic behavior assumed for $c_{l}$, we can take this constant of motion proportional to $l^{4+2\epsilon}$ at dominant order for large $l$, with $\epsilon>0$, and with proportionality constant denoted as $C$, so that $c_{[l]_{\text{ent}}}\sim C y^{2+\epsilon}$. Consequently, we conclude the convergence of Eq. \eqref{FirstSum}, because the sum over $n$ and $y$ of any sequence of fractions of the form $1/[P n^2 + Q y^{2+\epsilon}]$ (with $P$ and $Q$ being positive numbers independent of $n$ and $y$) converges if $\epsilon>0$. In our case, we can simply make $P=4\pi^2$ and $Q=C L_0^2$.

Let us consider now the second term in inequality \eqref{2Sums}. Apart from a factor that is bounded at  each instant with respect to its dependence on the mode labels, and disregarding subdominant contributions that do not modify the square summability in norm, this summability demands the convergence of
\begin{equation}\label{ssc}
    \sum_{nl} \frac{l^5}{(4\pi^2 n^2+C\,L_0^2  l^{4+2\epsilon})(4\pi^2 n^2+ L_0^2 l^2)},
\end{equation}
where we have used again $c_{l}\sim C l^{4+2\epsilon}$. Note that the first factor in the denominator is smaller than its value at vanishing $n$, and therefore the summand is bounded above by $l^{1-2\epsilon}/[C\,L_0^2 (4\pi^2 n^2+ L_0^2 l^2)]$. But $n^2 l^{2\epsilon-1}$ is larger than $n^2$ for $l>1$ if $\epsilon>1/2$, and then the sum \eqref{ssc} is bounded from above by the sum of $1/[C\,L_0^2 (4\pi^2 n^2+ L_0^2 l^{1+2\epsilon})]$ over $n$ and $l$. Clearly, this last sum converges for $\epsilon>1/2$. In summary, we have proved that the second term in expression \eqref{2Sums} is square summable if $c_{l}$ grows as a power greater than 5 for large $l$. In other words, convergence follows if $c_{l}$ asymptotically dominates $l^5$ as a higher power. We emphasize that this condition is satisfied, e.g., by the Chandrasekhar transformation, for which the sequence of constants $c_{l}$ is of order $l^8$\cite{GRG}.

This concludes the proof of unitary implementation of the typical DTs in the privileged massless Fock representation selected by the criterion of symmetry invariance and existence of unitarily implementable dynamics. 

\subsection{Massless representations}

In the above analysis, we have made use of the massless Fock representation of the oscillator-adapted $\GS$ variables. The studies about the determination of a preferred family of unitarily equivalent Fock quantizations presented in Refs. \cite{GnR,Beatles} apply straightforwardly to these variables owing to the properties of their Hamiltonian, which generates dynamics admitting a unitary implementation in a unique family of equivalent representations containing the aforementioned massless one. However, this massless representation is different from the massless representation associated with the CPM variables. In fact, we will show that the map between the two types of variables (i.e. the oscillator-adapted $\GS$ variables and the CPM variables), given in terms of the canonical basis by $\mathscr{S}\,\mathscr{R}_{\l}^{-1}$, cannot be unitarily implemented in our privileged family of Fock quantizations. Therefore, the two massless representations do not fall in the same unitary class. The proper choice of quantization for the CPM variables must thus be obtained by adopting the massless representation in the oscillator-adapted $\GS$ description and changing variables from it. 

Actually, we can go a step further and demonstrate that not even representations with a constant mass $M$ for the CPM variables can be unitarily equivalent to the massless representation of the $\GS$ description.

Let then $w_{k}^{(M)}=k\sqrt{1+M^2/k^2}=k+\vartheta_{k}$ be the corresponding mode frequency. The real subleading term $\vartheta_{k}$ is then $\mathcal{O}(k^{-1})$. Introducing the proper transformations between the canonical and the creation and annihilation variables, we obtain the following Bogoliubov transformation:
\begin{equation}\label{massrep}
    \begin{pmatrix}
        a_{\n}\\
        a^{*}_{\n}
    \end{pmatrix}_{\M} = \mathscr{F}^{\vartheta\,}_{nl} \;\mathscr{S_{\,}}\;\mathscr{R}_{\l}^{-1}\left( \tilde{\mathscr{F}}^{\mathrm{o}}_{nl}\right)^{-1}
    \begin{pmatrix}
        a_{\n}\\
        a^{*}_{\n}
    \end{pmatrix},
\end{equation}
where the matrix $\tilde{\mathscr{F}}^{\mathrm{o}}_{nl}$ and the creation and annihilation variables on the right-hand side were defined in Eq. \eqref{reprensation}, and
\begin{equation}
    \mathscr{F}^{\vartheta}_{nl}= \frac{1}{\sqrt{2(k+\vartheta_{k})}}\begin{pmatrix}
        k+\vartheta_{k} & i\\
        k+\vartheta_{k} & -i
    \end{pmatrix}.
\end{equation}
The variables on the left-hand side of 
Eq. \eqref{massrep} then correspond to the constant-mass representation for the CPM variables. 

The Hilbert-Schmidt condition on the antilinear part of the Bogoliubov transformation \eqref{massrep} is then the sufficient and necessary condition for the unitary equivalence of the analyzed Fock representations. This requires the summability over all modes of the square of the complex norm of the corresponding coefficients $\beta_{nl}^{\vartheta}$: 
\begin{equation}
    {\beta}_{nl}^{\vartheta} = \frac{\bs^{1/2}}{2} \sqrt{\frac{k}{k+\vartheta_{k}}} - \frac{\bs^{-1/2}}{2}\sqrt{\frac{k+\vartheta_k}{k}} -i \, \frac{\mathcal{G}_{\l}}{2\sqrt{k(k+\vartheta_{k})}}.
\end{equation}
Consequently, both the real and imaginary parts of these coefficients must be square summable. However, the square norm of the real part is
\begin{equation}\label{rep}
    \frac{1}{4}\left|\bs^{1/2} \left(1+\frac{\vartheta_{k}}{k} \right)^{-1/2} - \bs^{-1/2} \left( 1+\frac{\vartheta_{k}}{k}\right)^{1/2} \right|^2 = \frac{\frac{\vartheta_{k}^2}{k^2} + 2\frac{\vartheta_{k}}{k}(1-\bs)+(1-\bs)^2}{4\bs \left(1+ \frac{\vartheta_{k}}{k} \right)}.
\end{equation}
Both the numerator and denominator are of asymptotic order unity for large values of $k$. It is then simple to show that the studied sequence cannot be square summable, and hence that the two representations that we are considering are inequivalent, for any constant value of the mass $M$. 

This result corrects a statement in Ref. \cite{Master}, making clear that the correct Fock quantization for the description in terms of $\M$ variables is the one obtained from the massless representation of the oscillator-adapted $\GS$ formulation by a change of variables. 

\section{Unsaturated freedom in quantizations with unitary dynamics}\label{Unsaturated}

We have proved that the canonical realization of DTs guarantees unitary implementability in the family of representations selected by the criterion of symmetry invariance and unitary dynamics of Ref. \cite{GnR}. This means, in particular, that DTs in this Hamiltonian framework generate new representations that remain within the selected family\footnote{In this regard, we also note that DTs respect the invariance under background symmetries.}. A natural question subsequently emerges regarding the completeness of this approach, namely whether every representation within the family admitting unitary dynamics lies in fact in the image of a DT, or does the family of preferred representations possess additional choices, and hence unsaturated freedom. As we demonstrate below, the answer to this question is in the negative: the class of representations generated by DTs does not cover all the freedom available for descriptions with unitary dynamics. 

To show this incompleteness, we present a simple counterexample of a canonical transformation (which we denote by $\mathscr{T}_{nl}$) that does not belong to the set of DTs, but is nevertheless unitarily implementable with respect to the massless representation of the oscillator-adapted $\GS$ variables. Since the massless representation itself supports unitary dynamics, this transformation generates another representation with unitary dynamics. Like DTs, this transformation $\mathscr{T}_{nl}$ preserves the spatial symmetry structure of the massless representation, and therefore the new representation belongs to the preferred family of representations, though not being reachable by means of a DT.

In the oscillator-adapted $\GS$ variables, the transformation $\mathscr{T}_{nl}$ is given by
\begin{equation}
    \begin{pmatrix}
        \tilde{\phi}_{\n}^{\GS} \\
        \tilde{\Pi}_{\n}^{\GS}
    \end{pmatrix}_\mathscr{T} = \mathscr{T}_{nl} \begin{pmatrix}
        \tilde{\phi}_{\n}^{\GS} \\
        \tilde{\Pi}_{\n}^{\GS}
    \end{pmatrix}, \; \, \text{with} \; \; \mathscr{T}_{nl}= \begin{pmatrix}
        \sigma_{nl} & 0 \\
        0 & \sigma_{nl}^{-1}
    \end{pmatrix},
\end{equation}
where $\sigma_{nl}$ depends on the labels $n$ and $l$, but it is independent of $\tau_{\star}$ (or $\tau$) for simplicity. Clearly, this transformation is not of the DT type, expressed e.g.~in Eq. \eqref{Bogoliu}. Furthermore, let us take $\sigma_{nl}$ to be of the form  $\sigma_{nl}=1+\Upsilon_{nl}k^{-\varepsilon}$, where $\varepsilon$ is a (nondynamical) parameter that we choose greater than $3/2$, and the coefficients $\Upsilon_{nl}$ are such that $|\Upsilon_{nl}|\leq 1$ for all modes.
	
Starting from the massless representation, let us now show that $\mathscr{T}_{nl}$ generates a unitarily equivalent representation or, which is to say the same, is unitarily implementable. Like in Sec. \ref{UniD}, this amounts to the square summability of the coefficients $\mathring{\beta}_{nl}$ of the antilinear part of the Bogoliubov transformation determined by $\tilde{\mathscr{F}}^{o}_{nl}\mathscr{T}_{nl}(\tilde{\mathscr{F}}^{o}_{nl})^{-1}$. In the present case, we find simply that
\begin{equation}\label{circsum}
     \left| \mathring{\beta}_{nl}\right|^2 =\frac{1}{4} \left|\sigma_{nl} -\frac{1}{\sigma_{nl}} \right|^2.
 \end{equation} 
To discuss the square summability over all modes, we consider the dominant order of Eq. \eqref{circsum} in the ultraviolet sector, which is given by $|\Upsilon_{nl}|^2 k^{-2\varepsilon}$. Hence, taking into account the upper bound on $|\Upsilon_{nl}|$, that the sum over the label $m$ produces a factor $2l+1$, and recalling that $k^2= (4\pi^2/L_0^2) n^2 + l(l+1)$, the summability of the sequence of terms displayed in the above equation simply reduces to the convergence of 
\begin{equation} 
2L_0^{2\varepsilon} \, \sum_{nl} \frac{l}{\left(4\pi^2 n^2+L_0^2 l^2\right) ^{\varepsilon}}.
\end{equation} 
Defining $z^2=4\pi^2 n^2+L_0^2 l^2$, we have  $l\leq z/L_0$, (with $z>0$), and also that the number of possible integer values of $l$ for fixed $z$ is bounded by $2\pi z/L_0$ in the ultraviolet (since $2\pi z$ is the length of the circle with fixed radius $z$ in the bidimensional plane of labels coordinatized by $2\pi n$ and $L_0 l$, extended to real values). Therefore, the above sum converges if $4\pi L_0^{2\varepsilon-2}\sum_z z^{2(1-\varepsilon)}$ converges in the ultraviolet with our choice $\varepsilon > 3/2$, which is indeed the case\footnote{We notice that the density of allowed values of $z$ in the ultraviolet is bounded by $1/(2\pi + L_0)$.}. This concludes our proof.

\section{Conclusions}\label{Conclusions}

The description of DTs as a set of canonical transformations in the Hamiltonian framework of nonrotating BH geometric perturbations opens a new avenue for investigating how a classical symmetry affects the quantum theory, concretely its representation, serving as a particular example. In the present work, we have taken the BH interior as our starting point. Since all elements in our analysis, including the gauge invariant perturbations, are expressed in terms of phase space variables, the extension of the quantum field theory to the exterior can be accomplished by introducing a symplectomorphism. Under the transformation
\begin{equation}
p_b \to i \, p_b, \qquad b\to -i b, \qquad p_c \to p_c , \qquad c \to c,
\end{equation} 
the roles of the coordinates in Eq. \eqref{1.1} interchange. Since the variable $\Omega_b$ is proportional to the product of the two canonical background variables of the $b$-sector, and the Hamiltonian (of the background and the gauge invariant perturbations) depends on this sector only through $\Omega_b$ and even powers of $p_b$, the reality of these functions on phase space is preserved. For all Hamiltonian considerations, the complex symplectomorphism merely changes the domain of definition of $p_b^2$ from positive to negative values. 

On the other hand, by adopting the so-called oscillator-adapted $\GS$ variables, we were able to select a (unique) representation on the interior of the BH (when the background is classically or effectively described). This representation belongs to a family of equivalent representations in which the dynamics can be made unitarily implementable. We have then introduced a general canonical transformation to reach the so-called CPM variables, on which the DTs act naturally. We have demonstrated that, in the above unique family of Fock representations, the canonical transformations associated with this class of DTs are unitarily implementable under very mild assumptions. More specifically, our proof applies to a general class of potentials $v_{l}$ that, similarly to what happens with the Regge-Wheeler potential $V_{l}^{\p}$, grow asymptotically as proportional to $l^2$ for large $l$, and differ from $V_{l}^{\p}$ by terms that are of the asymptotic order of $\partial_{\tau_{\star}}V_{l}^{\p}/l^5$ or negligibly smaller. This assumption covers a large set of potentials, including of course the Regge–Wheeler potential itself, and the Zerilli potential \cite{Gpap,Master,GRG}. Our asymptotic condition on the difference of potentials comes from the requirement that the constant of motion $c_{l}$ associated with the DT be asymptotically of order $l^5$ or of a higher power, a condition much weaker than the most important physical example, namely the Chandrasekhar transformation, where this constant scales as $l^8$ \cite{GRG}.

Although at certain stages we have referred to backgrounds in General Relativity to illustrate our discussion, all the calculations are valid for effective backgrounds if one evaluates the corresponding phase space variables and their dynamical derivatives on effective trajectories. This is achieved by computing Poisson brackets with the corresponding effective Hamiltonian. Such a Hamiltonian incorporates effective corrections that encode the influence of the physical phenomena departing from the classical Einsteinian theory. One interesting example is the effective description of loop quantum cosmology for BHs, for instance, for the effective geometries put forward in Refs. \cite{AOS,AOS2,AO} to take into account loop quantum gravity effects. In this sense, it is worth commenting that, whereas the Hamiltonian formalism in the interior may find obstructions in the essential BH singularity of classical General Relativity, requiring a careful analysis if it is not excluded from within the interval of dynamical evolution, this singularity is solved in effective descriptions such as those of loop quantum cosmology \cite{lqc}, allowing a well-defined Hamiltonian treatment in the whole interior region, extending beyond the classical singularity. Moreover, since the extension of the Hamiltonian treatment of the perturbations to the exterior is accomplished by a change in the domain of definition of certain functions on the $b$-sector of phase space, as far as this change does not modify any of the arguments presented in our proof of unitary implentability of the DTs, the results about such unitarity should still hold when conveniently applied to the extended quantum field theory of the gauge invariant perturbations. A careful discussion of this issue will be carried out elsewhere \cite{RoyelOtis}.

In a certain sense, our results about the unitarity of the DTs are stronger than the well-known results in General Relativity about the equivalence of the physics described by the master functions and master equations that these DTs relate, which hold for each fixed value of the angular momentum $l$. These last results include, in particular, the equivalence of the spectrum of quasinormal modes in the exterior geometry, i.e. their isospectrality. This stronger nature of our unitarity results does not only reflect the aforementioned applicability to effective backgrounds, but also the remarkable fact that they refer to the entire set of modes with all possible values of $l$, and not just to modes with a given value of this angular momentum label.  

By studying the canonical map from the oscillator-adapted $\GS$ variables to the CPM variables, we have also demonstrated that the choice of variables in which the representation is selected is important. As our proof shows, this map is not unitarily implementable by itself. So, for instance, making a massless representation choice in one or the other set of variables does not yield equivalent representations. Therefore, if unitarily implementable dynamics is sought, or used as the physical criterion to select a privileged quantization, the representation must be selected in the oscillator-adapted $\GS$ variables and transported via the canonical map to the CPM variables. 

Another goal of this article has been achieved by showing that, within our preferred family of Fock quantizations, there exist unitarily implementable transformations (respecting therefore the unitary dynamics) that cannot be generated by any canonical DT. Thus, the family of DTs does not exhaust all the freedom in the choice of representations within the family admitting unitarily implementable dynamics. 

Crucially, our findings provide a rigorous theoretical justification for works such as Ref. \cite{BlackPumas}, where a classical DT mapping between the axial and polar sectors was employed at the classical level without explicitly verifying its potential quantum implications. By establishing the unitary implementability of these maps, our results confirm that the effective master Regge-Wheeler equation derived in that work via hybrid loop quantum cosmology \cite{hybrid} methods remains robust, unaffected by quantum modifications arising from the relation between the potentials.

This work paves the way for several future investigations. Among them is the possibility of exploring whether and how isospectrality can be extended to effective backgrounds. Having established that DTs are unitarily implementable in the representations considered here, and that this result is valid even for effective backgrounds provided that all phase variables and their derivatives are suitably evaluated on effective trajectories, it is natural to analyze in detail how this quantum-corrected effective description modifies the correspondence between Darboux-related potentials in General Relativity, and how isospectrality may be broken if such modifications are not incorporated \cite{CorralOlmedo,Leno}.

Finally, it would be worthwhile to investigate whether the procedure to select a unique representation that we have employed here can instead be formulated in terms of local geometric and canonical data, rather than relying on asymptotic properties of its dynamics in a suitably chosen description. For example, Ref. \cite{GiddingsPerk} explores criteria for characterizing physically appropriate representations in curved spacetimes directly from local data. Within the present framework, it would be interesting to investigate how these proposals transform under DTs, and whether they can be used to select the appropriate representation directly in the CPM variables. 

\acknowledgments

We are grateful to Beatriz Elizaga Navascués and Andrés Mínguez-Sánchez for discussions. A.T.-C. acknowledges financial support from Comunidad Autónoma de Madrid through the PIPF-2024 fellowship (Ref. PIPF-2024/TEC-34465). G.A.M.M. and A.T.-C. acknowledge support from MCIN/AEI/10.13019/501100011033 and FSE+ under the Grant No. PID2023-149018NB-C41 of the Spanish Ministry of Science and Innovation. J.V.\ is grateful for the support given by the research unit Fiber Materials and Environmental Technologies (FibEnTech-UBI), on the extent of the project reference UIDB/00195/2020, funded by the Funda\c c\~ao para a Ci\^encia e a Tecnologia (FCT), IP/MCTES through national funds (PIDDAC).


\begin{thebibliography}{99}

\bibitem{Singularity} J.D. Barrow and S.Z.W. Lip, The classical stability of sudden and Big Rip singularities, Phys. Rev. D {\bf 80}, 043518 (2009).

\bibitem{Nollert} H.-P. Nollert, Quasinormal modes: The characteristic “sound” of black holes and neutron stars, Class. Quantum Grav. {\bf 16},
R159 (1999).

\bibitem{Berti} E. Berti et al., Black hole spectroscopy: From theory to experiment, Class. Quantum Grav. {\bf 43}, 123001 (2026).

\bibitem{LISA} M. Piarulli, S. Marsat, E.M. Sänger, A. Buonanno, J. Steinhoff, and N. Tamanini, Parametrized test of general relativity for LISA massive black hole binary inspirals, Phys. Rev. D {\bf 112}, 124044 (2025).

\bibitem{ET} A. Begnoni, W. Del Pozzo, M. Pegorin, J. Pomper, and A. Ricciardone, Tests of general relativity with Einstein Telescope, arXiv:2511.07520 (2025).

\bibitem{Tests} E. Berti et al., Testing general relativity with present and future astrophysical observations, Class. Quantum Grav. {\bf 32}, 243001 (2015).

\bibitem{Barack} L. Barack et al., Black holes, gravitational waves and fundamental physics: A roadmap,  Class. Quantum Grav. {\bf 36}, 143001 (2019).

\bibitem{Frolov} V. Frolov and I. Novikov,  {\it Black Hole Physics: Basic Concepts and New Developments} (Springer, Dordrecht, The Netherlands, 1998).

\bibitem{TThiemann} T. Thiemann, Quantum field theory of black hole perturbations with backreaction: I General framework, Universe {\bf 10}, 372 (2024).

\bibitem{ReggeWheeler} T. Regge and J.A. Wheeler, Stability of a Schwarzschild singularity, Phys. Rev. {\bf 108}, 1063 (1957).

\bibitem{Bardeen} J.M. Bardeen, Gauge-invariant cosmological perturbations, Phys. Rev. D {\bf 22}, 1882 (1983).

\bibitem{Mukhanov} V. Mukhanov, Quantum theory of gauge invariant cosmological perturbations, Zh. Eksp. Teor. Fiz. {\bf 94}, 1 (1988) [J. Exp. Theor. Phys. {\bf 67}, 1297 (1988)].

\bibitem{Sasaki} M. Sasaki, Gauge invariant scalar perturbations in the new inflationary universe, Prog. Theor. Phys. {\bf 70}, 394 (1983).

\bibitem{Gpap} M. Lenzi, G.A. Mena Marugán, A. Mínguez-Sánchez, and C.F. Sopuerta, Master variables and Darboux symmetry for axial perturbations of the exterior and interior of black hole spacetimes, Phys. Rev. D {\bf 113}, 064039 (2026).

\bibitem{Master} M. Lenzi, G.A. Mena Marug{\'a}n, and A. M{\'\i}nguez-S{\'a}nchez, Master functions and hybrid quantization of perturbed nonrotating black hole interiors, Class. Quantum Grav. {\bf 43}, 105007 (2026).

\bibitem{GRG} M. Lenzi, G.A. Mena Marug{\'a}n, A. M{\'\i}nguez-S{\'a}nchez, and C. F. Sopuerta, Hamiltonian formalism, master functions and Darboux transformations for perturbed (interiors and exteriors of) nonrotating black holes, Gen. Relativ. Gravit. {\bf 58}, 56 (2026).

\bibitem{CPM} C.T. Cunningham, R.H. Price, and V. Moncrief, Radiation from collapsing relativistic stars. I Linearized odd-parity radiation, Astrophys. J. {\bf 224}, 643 (1978).

\bibitem{Ze} F.J. Zerilli, Effective potential for even parity Regge Wheeler gravitational perturbation equations, Phys. Rev. Lett. {\bf 24}, 737 (1970).

\bibitem{Mo} V. Moncrief, Gravitational perturbations of spherically symmetric systems. I. The exterior problem, Ann. Phys. (N.Y.) {\bf 88}, 323 (1974).

\bibitem{Darboux} G. Darboux, Sur une proposition relative aux \'equations lin\'eaires, C. R. Acad. Sci. (Paris) {\bf 94}, 1456 (1882).

\bibitem{Chandra} S. Chandrasekhar, {\it The Mathematical Theory of Black Holes} (Oxford University Press, Oxford, England, 1985).

\bibitem{Chandra1} S. Chandrasekhar, On the equations governing the perturbations of the Schwarzschild black hole, Proc. Roy. Soc. Lond. A {\bf 343}, 289 (1975).

\bibitem{Chandra2} S. Chandrasekhar and S.L. Detweiler, The quasi-normal modes of the Schwarzschild black hole, Proc. Roy. Soc. Lond. A {\bf 344}, 441 (1975).

\bibitem{Chandra3} S. Chandrasekar, On one-dimensional potential barriers having equal reflexion and transmission coefficients, Proc. Roy. Soc. Lond. A {\bf 369}, 425 (1980).

\bibitem{Yurov} A. Yurov and V. Yurov, A look at the generalized Darboux transformations for the quasinormal spectra in Schwarzschild black hole perturbation theory: Just how general should it be?, Phys. Lett. A {\bf 383}, 2571 (2019).

\bibitem{Glampe} K. Glampedakis, A.D. Johnson, and D. Kennefick, The Darboux transformation in black hole perturbation theory, Phys. Rev. D {\bf 96}, 024036 (2017).

\bibitem{CorralOlmedo} D. del-Corral and J. Olmedo, Breaking of isospectrality of quasinormal modes in nonrotating loop quantum gravity black holes, Phys. Rev. D {\bf 105}, 064053 (2022).

\bibitem{Leno} A. García-Quismondo, G.A. Mena Marugán, and A. Torres-Caballeros, Quasinormal modes in effective loop quantum gravity and consequences for isospectrality, arXiv:2608.11088 (2026).

\bibitem{AOS} A. Ashtekar, J. Olmedo, and P. Singh, Quantum transfiguration of Kruskal black holes, Phys. Rev. Lett. {\bf 121}, 241301 (2018).

\bibitem{AOS2} A. Ashtekar, J. Olmedo, and P. Singh, Quantum extension of the Kruskal spacetime, Phys. Rev. D {\bf 98}, 126003 (2018).

\bibitem{AO} A. Ashtekar and J. Olmedo,
Properties of a recent quantum extension of the Kruskal geometry, Int. J. Mod. Phys. D {\bf 29}, 2050076 (2020).

\bibitem{CM1} M. Lenzi and C.F. Sopuerta, Darboux covariance: A hidden symmetry of perturbed Schwarzschild black holes, Phys. Rev. D {\bf 104}, 124068 (2021).

\bibitem{CM2} M. Lenzi and C.F. Sopuerta, Master functions and equations for perturbations of vacuum spherically symmetric spacetimes, Phys. Rev. D {\bf 104}, 084053 (2021).

\bibitem{PRD83} J. Cortez, G.A. Mena Marug\'an, J. Olmedo, and J.M. Velhinho, Uniqueness of the Fock quantization of fields with unitary dynamics in nonstationary spacetimes, Phys. Rev. D {\bf 83}, 025002 (2011).

\bibitem{DreamTeam} J. Cortez, B. Elizaga Navascu\'es, M. Mart\'{\i}n-Benito, G.A. Mena Marug\'an, J. Olmedo, and J.M. Velhinho, Uniqueness of the Fock quantization of scalar fields in a Bianchi I cosmology with unitary dynamics, Phys. Rev. D  {\bf 94}, 105019 (2016).

\bibitem{GnR} J. Cortez, B. Elizaga Navascués, G.A. Mena Marugán, A. Torres-Caballeros, and J.M. Velhinho, Fock quantization of a Klein-Gordon field in the interior geometry of a nonrotating black hole, Mathematics {\bf 11}, 3922 (2023).	

\bibitem{Beatles} J. Cortez, G.A. Mena Marug\'an, A. Torres-Caballeros, and J.M. Velhinho, Time-dependent scalings and Fock quantization of a massless scalar field in Kantowski-Sachs spacetime, Class. Quantum Grav. {\bf 41}, 195002 (2024).

\bibitem{ThePolice} J. Louko, G.A. Mena Marugán, and A. Torres-Caballeros, Expansion-sensitive coupling of a local quantum system in de Sitter cosmology, Phys. Rev. D {\bf 12}, 125024 (2025).

\bibitem{KS} R. Kantowski and R.K. Sachs, Some spatially inhomogeneous dust models, J. Math. Phys. {\bf 7}, 443 (1966).

\bibitem{Weber} E. Weber, Kantowski-Sachs cosmological models as big-bang models, J. Math. Phys. {\bf 26}, 1308 (1984).

\bibitem{Axial} G.A. Mena Marugán and A. Mínguez-Sánchez, Axial perturbations in Kantowski-Sachs spacetimes and hybrid quantum cosmology, Phys. Rev. D {\bf 109}, 106009 (2024). 

\bibitem{GS1} U.H. Gerlach and U.K. Sengupta, Gauge invariant perturbations on most general spherically symmetric space-times, Phys. Rev. D {\bf 19}, 2268 (1979).

\bibitem{GS2} U.H. Gerlach and U.K. Sengupta, Gauge invariant coupled gravitational, acoustical, and electromagnetic modes on most general spherical space-times, Phys. Rev. D {\bf 22}, 1300 (1980).

\bibitem{Sch} V. Ovsienko and S. Tabachnikov, What is …the Schwarzian derivative?, Not. Am. Math. Soc. {\bf 56}, 34 (2009).

\bibitem{BlackPumas} B. Elizaga Navascués and A. Torres-Caballeros, Effective Regge–Wheeler equations of a hybrid loop quantum black hole, Phys. Rev. D {\bf 113}, 124031 (2026).

\bibitem{Intertwining1} A.A. Suzko and A. Schulze-Halberg, Intertwining operator method and supersymmetry for effective mass Schrödinger equations, Phys. Lett. A. {\bf 372}, 5865 (2008).

\bibitem{Intertwining2} A.A. Suzko and A. Schulze-Halberg, Darboux transformations and supersymmetry for the generalized Schrödinger equations in (1+1) dimensions, J. Phys. A {\bf 42}, 295203 (2009).

\bibitem{Wald} R.M. Wald, {\it Quantum Field Theory in Curved Spacetime and Black Hole Thermodynamics} (University of Chicago Press, Chicago, USA, 1994).

\bibitem{Wald2} R.M. Wald, {\it The Formulation of Quantum Field Theory in Curved Spacetime} (Springer, New York, USA, 2018).

\bibitem{Winitzki} V. Mukhanov and S. Winitzki, {\it Introduction to Quantum Effects in Gravity} (Cambridge University Press, Cambridge, England, 2007).

\bibitem{AshtekarMagnon} A. Ashtekar and A. Magnon, Quantum fields in curved space-times, Proc. Roy. Soc. Lond. A {\bf 346}, 375 (1975).

\bibitem{Shale} D. Shale,  Linear symmetries of free boson fields, Trans. Am. Math. Soc. {\bf 103}, 149 (1962).

\bibitem{HR} R. Honegger and A. Rieckers, Squeezing Bogoliubov transformations on the infinite mode CCR‐algebra, J. Math. Phys. {\bf 37}, 4292 (1996).

\bibitem{lqc} A. Ashtekar and P. Singh, Loop quantum cosmology: A status report, Class. Quantum Grav. {\bf 28}, 213001 (2011).

\bibitem{RoyelOtis} B. Elizaga Navascu{\'e}s and A. Torres-Caballeros, private communication.

\bibitem{hybrid} B. Elizaga Navascu{\'e}s and G. A. Mena Marug{\'a}n, Hybrid loop quantum cosmology: An overview, Front. Astron. Space Sci. {\bf 8}, 81 (2021).

\bibitem{GiddingsPerk} S.B. Giddings and J. Perkins, Challenges for describing unitary evolution in nontrivial geometries: Pictures and representations,
arXiv:2507.13351 (2025).


\end{thebibliography}
\end{document}